\documentstyle[12pt]{article}

\begin{document}
\begin{center}
{\large ON A HARMONIC PROPERTY OF THE EINSTEIN MANIFOLD CURVATURE}\\
\vskip 0.4cm
BABOUROVA\enspace O.V. and FROLOV\enspace B.N. \\
\vskip 0.4cm
{\it Department of Mathematics, Moscow State Pedagogical University,\\
     Krasnoprudnaya 14, Moscow 107140, Russia;\\
     E-mail: baburova.physics@mpgu.msk.su,\enspace
frolovbn.physics@mpgu.msk.su}
\vskip 1cm
\end{center}

\begin{abstract}
{\small
\par
        The harmonicity condition of the curvature 2-form of a pseudo-
Riemannian manifold is formulated on the basis of annulment of this form
by the de Rham-Lichnerowicz Laplacian. The following  theorem is proved:
The curvature 2-form of any Einstein manifold is harmonic.

}
\end{abstract}

\vskip 0.2cm
\par
        It is well known [1] that on the metric space-time manifold
${\cal M}$ the Maxwell equations can be expressed as
\begin{equation}
dF = 0\;,\;\;\;\;\; \delta F=0\;, \label{eq:1}
\end{equation}
where $F$ is electromagnatic field strength 2-form and $d$, $\delta$ are
the operations of exterior differentiation and codifferentiation, respectively.
One has $\delta = *d*$ for even $n=$dim$ {\cal M}$ with additional factor (-1)
that occurs for space with negative signature. On the manifold ${\cal M}$
the Hodge-de Rham Laplacian $\Delta^{(HR)} = d \delta + \delta d$ can be
constructed. The Maxwell equations (\ref{eq:1}) yield $\Delta^{(HR)}F = 0$,
so the Maxwell 2-form F is {\em harmonic} [1]. The question is: whether the
same result is valid for General Relativity, and if "yes" then in what sense.
\par
        On any Riemannian (pseudo-Riemannian) manifold one has the
tensor-valued curvature 2-form
\begin{equation}
\Omega = \frac{1}{2}(R^{\lambda}\!_{\sigma\mu\nu}e_{\lambda}\otimes
e^{\sigma})\otimes\theta^{\mu}\wedge\theta^{\nu}  \label{eq:2}
\end{equation}
and the operation of covariant exterior differentiation ${\cal D}$ [2].
One can introduce the generalized operation of the
covariant exterior codifferentiation as ${\cal D}^{*} = *{\cal D}*$ and
{\em the de Rham-Lichnerowicz Laplacian}
\begin{equation}
\Delta^{(RL)} = {\cal D}{\cal D}^{*} + {\cal D}^{*}{\cal D}\;. \label{eq:4}
\end{equation}
The operators  ${\cal D}$, ${\cal D}^{*}$ and $\Delta^{(RL)}$ act not only
on usual {\bf R}-valued forms but also on {\bf V}-valued forms (with values
in a linear vector space {\bf V}).
\par
        We shall denote Ricci tensor components in local coordinate frames as $
R_{\mu\nu} = R^{\lambda}\!_{\mu\lambda\nu}$. Then its trace $R =
R^{\nu}\!_{\nu}$
is the curvature scalar. A Riemannian (pseudo-Riemannian) manifold with Ricci
tensor satisfying the condition
\begin{equation}
R_{\mu\nu} = \Lambda g_{\mu\nu}\;,\qquad \Lambda = const \label{eq:5}
\end{equation}
is called an Einstein manifold [3].
\par
        {\bf Definition}: The curvature 2-form of Riemannian
(pseudo-Riemannian)
manifold is called {\em harmonic}, if one has
\begin{equation}
 \Delta^{(RL)}\Omega = 0\;. \label{eq:6}
\end{equation}
\par
        We shall prove the following theorem.
 \par
        {\bf Theorem:} {\em The curvature 2-form of any Einstein manifold
is harmonic}.
\par    The proof of the theorem will be realized in a local coordinate
frame $\theta^{\mu} = dx^{\mu}$. By generalizing the method developed in
[4] to any (1,1)-tensor-valued 2-form $\Omega$ one has
\begin{eqnarray}
{\cal D}\Omega = \frac{1}{3!}(3 \Omega^{\lambda}\!_{\sigma\mu\nu;\rho}
e_{\lambda}\otimes e^{\sigma})\otimes\theta^{\rho} \wedge\theta^{\mu}
\wedge\theta^{\nu}\;, \label{eq:7} \\
{\cal D}^{*}\Omega = (\Omega^{\lambda}\!_{\sigma\rho\nu}\!^{;\rho}
e_{\lambda}\otimes e^{\sigma})\otimes\theta^{\nu}\;,  \label{eq:8} \\
\Delta^{(RL)}\Omega = \frac{1}{2}(\Omega^{\lambda}\!_{\sigma\mu\nu;\rho}\!
^{;\rho}+ 2\Omega^{\lambda}\!_{\sigma\mu}\!^{\rho}\!_{;[\rho;\nu]}
- 2\Omega^{\lambda}\!_{\sigma\nu}\!^{\rho}\!_{;[\rho;\mu]})e_{\lambda}\otimes
e^{\sigma}\otimes\theta^{\mu}\wedge\theta^{\nu}\;. \label{eq:9}
\end{eqnarray}
Here $\Omega^{\lambda}\!_{\sigma\mu\nu;\rho}$ denotes the covariant derivative
components with respect to all idices $\lambda,\;\sigma,\;\mu,\;\nu$.
Applying now equation (\ref{eq:9}) to the curvature 2-form (\ref{eq:2}) and
using the Ricci identity, one gets
\begin{eqnarray}
\Delta^{(RL)}\Omega = \frac{1}{2}(\Delta^{(RL)}\Omega)^{\lambda}\!
_{\sigma\mu\nu} e_{\lambda}\otimes e^{\sigma}\otimes\theta^{\mu}
\wedge\theta^{\nu} \; , \label{eq:10} \\
(\Delta^{(RL)}\Omega)^{\lambda}\!_{\sigma\mu\nu} =
(R^{\lambda}\!_{\sigma\mu\nu ;\rho}\!^{;\rho} + R^{\lambda}\!_{\sigma
\nu}\!^{\alpha} R_{\alpha\mu} - R^{\lambda}\!_{\sigma\mu}\!^{\alpha}
R_{\alpha\nu} \nonumber \\
- 2R^{\lambda}\!_{\sigma\rho}\!^{\alpha} R_{\alpha\nu\mu}\!^{\rho}
+ 2R^{\lambda}\!_{\alpha\mu}\!^{\rho} R^{\alpha}\!_{\sigma\rho\nu}
- 2R^{\lambda}\!_{\alpha\nu}\!^{\rho} R^{\alpha}\!_{\sigma\rho\mu})
\; . \label{eq:11}
\end{eqnarray}
\par
        On the other hand, by covariant differentiating  the Bianchi identity
with the following contraction  one gets
\begin{equation}
R^{\lambda}\!_{\sigma\mu\nu ;\rho}\!^{; \rho} = R^{\lambda}\!
_{\sigma\mu}\!^{\rho}\!_{;\nu ;\rho} - R^{\lambda}\!_{\sigma\nu}\!
^{\rho}\!_{;\mu ;\rho} \; . \label{eq:12}
\end{equation}
With the help of the Ricci identity (\ref{eq:12}) yields
\begin{eqnarray}
R^{\lambda}\!_{\sigma\mu\nu ;\rho}\!^{;\rho} = R^{\lambda}\!_{\sigma
\mu}\!^{\rho}\!_{;\rho ;\nu} - R^{\lambda}\!_{\sigma\nu}\!^{\rho}\!
_{;\rho ;\mu} + R^{\lambda}\!_{\sigma\mu}\!^{\alpha} R_{\alpha\nu} -
R^{\lambda}\!_{\sigma\nu}\!^{\alpha}R_{\alpha\mu} \nonumber \\
- 2R^{\lambda}\!_{\sigma\rho}\!^{\alpha} R_{\alpha\mu\nu}\!^{\rho}
+ 2R^{\lambda}\!_{\alpha\mu}\!^{\rho} R^{\alpha}\!_{\sigma\nu\rho}
- 2R^{\lambda}\!_{\alpha\nu}\!^{\rho} R^{\alpha}\!_{\sigma\mu\rho}
\; . \label{eq:13}
\end{eqnarray}
\par
        Now taking into account the following consequence of the Bianchi
identity
\begin{equation}
R_{\mu\nu\sigma}\!^{\rho}\!_{;\rho} = 2R_{\sigma[\mu ;\nu]}\; , \label{eq:14}
\end{equation}
let us substitute (\ref{eq:13}) into the expression (\ref{eq:11}). Then
we obtain the resultant expression for the action of the de Rham-Lichnerowicz
Laplacian to the pseudo-Riemannian curvature 2-form:
\begin{equation}
\Delta^{(RL)}\Omega = 2 R_{\mu[\lambda ;\sigma ]\nu} e^{\lambda}\otimes
e^{\sigma}\otimes\theta^{\mu}\wedge\theta^{\nu} \; . \label{eq:15}
\end{equation}
The expression (\ref{eq:15}) vanishes when the condition (\ref{eq:5}) is
fulfilled. Therefore the curvature 2-form is harmonic when the given
pseudo-Riemannian manifold is the Einsteinian one, as was to be proved.
\par
        It is easy to see that the converse statement is incorrect. There
exist curvature 2-forms that are harmonic, but not Einsteinian.
\par
        The theorem proved above shows that the field strength harmonicity
requirement is the reasonable condition also in the gravitatonal theory.
This fact can be applied to the problem of Lagrangian choice in modern theories
of gravitation, imposing the requirement of harmonicity on the corresponding
curvature 2-forms.
\vskip 0.6cm
{\bf References}
\vskip 0.4cm
\begin{description}
\item [{1.}]
T. Eguchi, P.B. Gilkey and A.J. Hanson,
{\em Physics Reports} {\bf 66} (1980) 213.
\item [{2.}]
R. Sulanke und P. Wintgen, {\em Differentialgeometrie und faserb\"{u}ndel}
(Hoch\-schul\-b\"{u}\-cher f\"{u}r mathematik, band 75) (VEB Deutscher Verlag
der
Wissenshaften, Berlin, 1972).
\item [{3.}]
A.L. Besse, {\em Einstein Manifolds} (Springer-Verlag, Berlin,
Heidelderg, 1987).
\item [{4.}]
G. de Rham, {\em Vari\'{e}t\'{e}s Diff\'{e}rentiables}
  (Hermann \& C$^{le}$, Paris, 1955).
\end{description}
\end{document}